\documentclass{optica-article}

\journal{opticajournal} 

\articletype{Research Article}

\usepackage{lineno}

\begin{document}

\title{Two-photon microscopy using picosecond pulses from four-wave mixing in a Yb-doped photonic crystal fiber}

\author{Bartosz Krawczyk,\authormark{1,*} Alexandre Kudlinski,\authormark{2} Robert T. Murray,\authormark{1} Simon R. Schultz,\authormark{3} Amanda J. Foust,\authormark{3} and Timothy H. Runcorn\authormark{1}}

\address{\authormark{1}Department of Physics, Imperial College London, Prince Consort Road, London SW7~2BW, UK\\

\authormark{2}Universit\'{e} de Lille, CNRS, UMR 8523-PhLAM–Physique des Lasers Atomes et Mol\'{e}cules, F-59000~Lille, France\\

\authormark{3}Department of Bioengineering, Imperial College London, London SW7~2AZ, UK\\}

\email{\authormark{*}bk3017@ic.ac.uk} 


\begin{abstract*} 
Two-photon microscopy (TPM) enables deep tissue imaging but requires excitation pulses that have a large product of average and peak power, typically supplied by femtosecond solid-state lasers. However, these lasers are bulky and femtosecond pulses require careful dispersion management to avoid pulse broadening, particularly when delivery fibers are used. Here we present a compact, fiber-based picosecond laser source operating at 790~nm for TPM using an ytterbium-doped photonic crystal fiber (Yb-doped PCF). The Yb-doped PCF simultaneously amplifies 1064~nm input pulses and efficiently converts them to 790~nm via four-wave mixing, generating pulses with a peak power of up to $\sim$3.8~kW. The source has a variable repetition rate (1.48~MHz--14.78~MHz), enabling the two-photon excitation fluorescence signal to be maximized in the presence of excitation saturation. We benchmark our picosecond laser source against a femtosecond Ti:Sapphire laser for TPM of stained \textit{Convallaria majalis} samples and demonstrate comparable fluorescence signal when the two-photon excitation conditions are matched.
\end{abstract*}


\section{Introduction}  
\label{sec:intro}  

Two-photon microscopy (TPM) has become a vital tool in the biological sciences due to its ability to image  deep into scattering tissue with intrinsic optical sectioning~\cite{denk1990two, hoover2013advances}. Two-photon absorption requires extremely high photon flux density, which is typically achieved by focusing ultrashort pulses with high peak power on the sample. Femtosecond Ti:Sapphire lasers are the industry standard source for TPM as they provide ultrashort ($<$200~fs) pulses with Watt-level average power across a wide tuning range (680--1080~nm), suitable for exciting the most commonly used fluorophores~\cite{Xu1996crosssections}. However, Ti:Sapphire lasers are typically very bulky and have a high cost of ownership, making deployment outside specialist laboratories challenging.

Fiber lasers are inherently compact, robust and maintenance-free, making them a compelling alternative to Ti:Sapphire lasers for portable and cost-effective TPM~\cite{xu2013recent}. However, commercial fiber lasers operating at 780~nm or 920~nm for TPM have largely concentrated on femtosecond pulse durations, utilizing bulk pulse compressors that negate some of these practical advantages. Femtosecond pulses require careful dispersion management to avoid temporal broadening reducing the peak power at the sample. This is particularly problematic when fiber delivery is required, e.g. for endoscopic TPM, typically necessitating the use of hollow-core fibers~\cite{septier2024hollow}.

Alternatively, it has long been known that picosecond pulses are effective for TPM~\cite{bewersdorf1998picosecond} and more recently picosecond fiber lasers have been demonstrated with comparable imaging quality to Ti:Sapphire lasers~\cite{kawakami2013picosecond,karpf2016two, kunio2024efficient}. This can be achieved by noting the time-averaged two-photon excitation fluorescence (TPEF) flux generated by picosecond pulses can match the flux generated by femtosecond pulses by adjusting the repetition rate and average power according to the following relationship~\cite{xu1996measurement}:  

\begin{equation}  
\label{equ:2p_abs}  
    F_{avg} \propto \frac{P_{avg}^2}{\tau f}, 
\end{equation}
where \( F_{avg} \) is the time-averaged TPEF flux in photon count per second, \( P_{avg} \) is the average power incident on the sample, \( \tau \) is the pulse duration and \( f \) is the repetition rate. Therefore, the potential reduction in TPEF flux generated using longer pulse durations can be compensated by reducing the repetition rate lower than the typical $\sim$80~MHz of Ti:Sapphire lasers.

Equation~(\ref{equ:2p_abs}), however, does not take into account excitation saturation~\cite{cianci2004saturation, rumi2010two}. In typical TPM experiments where \( f\lesssim\)200~MHz, the period between pulses is greater than the fluorescence lifetime of the fluorophore, hence saturation effects occur on a per-pulse basis. For a fixed average power, reducing the repetition rate of a pulse train to compensate for a longer pulse duration will increase the pulse energy of each pulse (\( E_{\mathrm{p}}=P_{avg}/f\)) and thus make excitation saturation more likely. To account for this, the relationship for the time-averaged TPEF flux generated with excitation saturation effects included is given by~\cite{zipfel2003nonlinear}:

\begin{equation}  
\label{equ:2p_abs_sat}  
    F_{avg} \propto f \left[1 - \exp\left(-\,a \frac{P_{avg}^2}{\tau f^2}\right) \right],
\end{equation}
where \(a\) is a constant that depends on the two-photon absorption cross section of the fluorophore, the excitation wavelength, and the focal spot size. For small pulse energies where saturation effects are negligible, the first-order Taylor expansion of Eq.~(\ref{equ:2p_abs_sat}) reduces to the relationship of Eq.~(\ref{equ:2p_abs}), as expected. In contrast to Eq.~(\ref{equ:2p_abs}), Eq.~(\ref{equ:2p_abs_sat}) predicts that \(F_{avg}\) does not increase indefinitely as the repetition rate is reduced. Instead, there exists an optimal repetition rate, \(f_{\mathrm{opt}}\), that maximizes \(F_{avg}\) for a given \(P_{avg}\) and \(\tau\) that from Eq.~(\ref{equ:2p_abs_sat}) scales as:
\begin{equation}
    \label{equ:opt_rr}
    f_{\mathrm{opt}} \propto \ \frac{P_{avg}}{\sqrt{\tau}}.
\end{equation}
For repetition rates above \(f_{\mathrm{opt}}\), the exponential term remains small and Eq.~(\ref{equ:2p_abs_sat}) approximates to Eq.~(\ref{equ:2p_abs}). Below \(f_{\mathrm{opt}}\), the TPEF flux generated per pulse does not continue to increase due to saturation, leading to \(F_{avg}\propto f\). Hence, for a given average power, there is a limit on how much a longer pulse duration can be compensated for by reducing the repetition rate to maintain sufficient TPEF flux. In addition, when excitation saturation occurs, the radial and axial effective point spread functions will broaden, resulting in decreased resolution~\cite{cianci2004saturation}.

In order to two-photon excite fluorophores around 800~nm, nonlinear wavelength conversion must be used to shift the output wavelength of Yb-doped fiber lasers. Four-wave mixing (FWM) in photonic crystal fibers (PCFs) has been demonstrated as an effective method for achieving the appropriate pulse parameters for TPM with the potential for full fiber-integration and wavelength tunability~\cite{lefrancois2012, gottschall2015fiber, gottschall2023ultrafast}. However, it remains a challenge to generate high peak power pulses using simple, compact systems due to limitations in the conversion efficiency of the FWM process.

Here, we present a picosecond fiber laser for TPM using a novel Yb-doped PCF that enhances the FWM conversion efficiency for low input peak powers. The Yb-doped PCF simultaneously amplifies the 1064~nm output from a fiber master oscillator power amplifier (MOPA) system and converts it to 790~nm via FWM. Picosecond ($\sim$12.5~ps) duration pulses at 790~nm were generated with average powers up to 150~mW and peak powers of up to $\sim$3.8~kW at repetition rates between 1.48~MHz--14.78~MHz. We test the laser against a femtosecond Ti:Sapphire laser using a commercial TPM system with stained \textit{Convallaria majalis} samples and demonstrate comparable imaging performance.

\section{Experimental setup}  
\label{sec:Set-up}  
\subsection{Fiber MOPA system}

\begin{figure}[htbp]
\centering\includegraphics[]{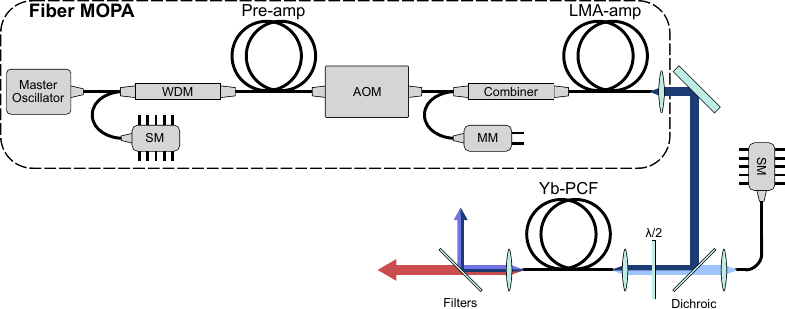}
\caption{Schematic of the FWM source, including the fiber MOPA system and ytterbium-doped photonic crystal fiber (Yb-PCF). SM, singlemode laser diode; MM, multimode laser diode; AOM, acousto-optic modulator; HWP, half-wave plate.}
\label{fig:set-up}
\end{figure}

Efficient FWM in PCFs requires input peak powers in the kilowatt range. To achieve this, we implemented a fiber master oscillator power amplifier (MOPA) system [Fig.~\ref{fig:set-up}]. The master oscillator was a mode-locked fiber laser operating at 1064~nm, delivering 15~ps pulses at a repetition rate of 29.55~MHz (MPB Communications EOAS-MLFL-P-15-30-1064-10). The master oscillator was amplified in two stages. The first stage was a 1~m long Yb-doped fiber pre-amplifier, core-pumped by a 976~nm single-mode (SM) laser diode with a maximum power of 780~mW. An acousto-optic modulator (AOM) was incorporated between the amplifier stages for pulse picking to adjust the repetition rate. After the AOM, the pulses were boosted by a large-mode area (LMA) amplifier utilizing a 1.1~m long double-clad 10/125~\textmu{}m Yb-doped fiber, cladding-pumped by a 3.5~W multimode (MM) 976~nm laser diode.

Operating the fiber MOPA system at lower repetition rates resulted in higher pulse energy and correspondingly higher peak power but at the expense of lower average power. The maximum repetition rate was limited by the response time of the pulse-picking electronics, while the minimum rate was constrained by the risk of Q-switching damage to the LMA amplifier. The fiber MOPA system could be operated at repetition rates from 14.78~MHz to 1.48~MHz, corresponding to peak powers of 3.3~kW to 16.6~kW at the PCF input, respectively.

\subsection{Ytterbium-doped photonic crystal fiber}  
\label{sec:pcf}

\begin{figure}[htbp]
\centering\includegraphics[width=\columnwidth]{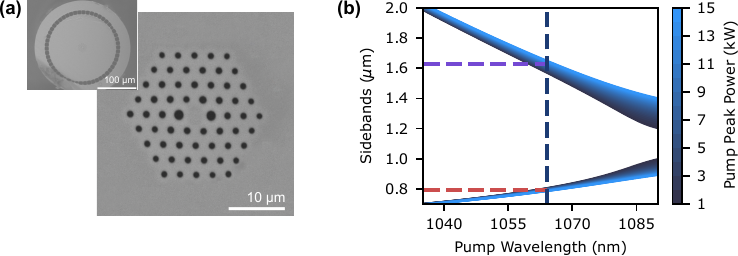}
\caption{(a) SEM image of the inner cladding and core of the Yb-doped PCF with a zoomed-out view showing the double-clad structure. (b) Phase-matching diagram for the fast axis of the PCF, indicating predicted anti-Stokes sideband at 790~nm (red) and Stokes sideband at 1630~nm (purple) for a 1064~nm FWM pump (dark blue).}
\label{fig:PCF}
\end{figure}

As shown in Fig.~\ref{fig:PCF}(a), the Yb-doped PCF used for FWM conversion featured a double-clad structure. While the outer cladding can be utilized for light collection in endoscopic applications~\cite{septier2022label}, this work focused exclusively on the inner cladding and core. The core included a 3~\textmu{}m diameter ytterbium-doped region. The inner cladding air-hole lattice had a pitch of 3.23~\textmu{}m and a hole diameter of 1.15~\textmu{}m. The Yb-doped PCF was polarization-maintaining due to the two larger air holes in the inner cladding, which induced birefringence~\cite{kudlinski2013simultaneous}. The axis along the two enlarged holes was the fast axis, while the slow axis was orthogonal. The linearly polarized FWM pump could be launched on either axis, enabling phase-matching tunability~\cite{harvey2003scalar, chen2005widely, murray2013widely}.

The dispersion of Yb-doped PCF was modeled using the Vector Effective Index Method~\cite{li2008improved}, which predicted a zero-dispersion wavelength of 1085~nm. Fig.~\ref{fig:PCF}(b) shows the modeled phase-matching diagram for the fast axis of the Yb-doped PCF. The predicted FWM sidebands were at 790~nm (anti-Stokes) and 1630~nm (Stokes) for a 1064~nm FWM pump. The FWM pump peak power affects phase-matching but has negligible impact for FWM pump wavelengths $<$1080~nm.

The modeled Yb-doped PCF dispersion predicted a walk-off length of $\sim$1~m for the 15~ps FWM pump pulses, suggesting that significant FWM would not occur beyond this length. Consequently, a 1~m-long Yb-doped PCF was used, which was end-capped with a flat input face and a $1.6^\circ$ angled output face, to reduce backreflections and suppress parasitic lasing.

As shown in Fig.~\ref{fig:set-up}, a SM continuous-wave (CW) 976~nm beam was combined with the FWM pump via a dichroic mirror and coupled into the Yb-doped PCF core using a free-space setup. This 976~nm pump, with maximum average power of 1.02~W at the input, created the population inversion required for amplification in the Yb-doped PCF. A half-wave plate (HWP) was used to align the FWM pump polarization to one of the principal axes of the Yb-doped PCF, ensuring optimal phase matching. The coupling efficiency into the Yb-doped PCF was estimated to be 75\% for both wavelengths. To isolate the anti-Stokes sideband for TPM, the Yb-doped PCF output was filtered to transmit only wavelengths between 685~nm and 900~nm.  

\subsection{Two-photon microscopy setup}  

\begin{figure}[htbp]
\centering\includegraphics[]{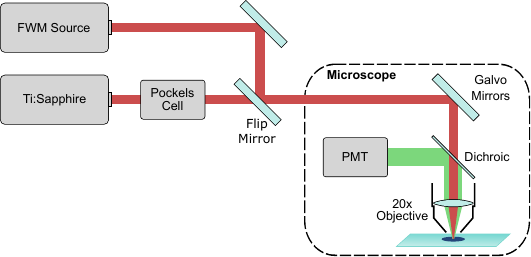}
\caption{Schematic of the two-photon microscopy setup. A flip mirror was used to switch between the Ti:Sapphire laser and the FWM source. Two galvo mirrors were used to scan the laser beam across the sample.}  
\label{fig:Microcope_Set-Up}  
\end{figure}

The generated anti-Stokes light from the Yb-doped PCF was directed into a galvo-scanning two-photon microscope (Scientifica MP-1000), designed for compatibility with a Ti:Sapphire laser (Spectra-Physics Mai Tai). To enable direct comparison between the two lasers, a flip mirror was installed in the Ti:Sapphire beam path [Fig.~\ref{fig:Microcope_Set-Up}]. To match the beam caustic of the Ti:Sapphire laser, several lenses were tested to collimate the Yb-doped PCF output. It was found an f$=$7.5~mm achromatic doublet (Thorlabs AC050-008-B-ML) gave the closest match and the collimation was fine-tuned to give the brightest two-photon excitation fluorescence (TPEF) images.

The laser beam was focused onto the sample using a 20x water-immersion objective (Olympus UMPLFLN20XW). The TPEF was collected through the same objective and directed to a photomultiplier tube (PMT) via a dichroic mirror. A bandpass filter placed in front of the PMT selected fluorescence in the red channel (590~nm--650~nm). Galvo mirrors raster-scanned the beam across the sample with a pixel dwell time of 3.2~\textmu{}s, generating 1024x1024 pixel images across a 605x605~\textmu{}m field of view, calibrated with a USAF target.

The Ti:Sapphire laser had a repetition rate of 80~MHz with a pulse duration of 132~fs, measured after a HWP and a polarizing beam splitter that was used for power control in conjunction with a Pockels cell. However, due to the absence of a pulse pre-chirper, the pulse duration increased to 263~fs at the sample from the dispersion of the Pockels cell and the microscope optics.

\section{Results}
\label{sec:results}

\subsection{FWM anti-Stokes generation}

\begin{figure}[htbp]  
\centering\includegraphics[]{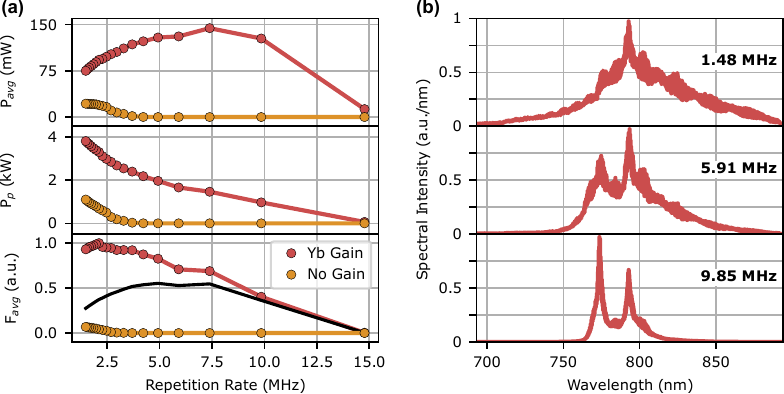}  
\caption{(a) Measured anti-Stokes average power (top), calculated peak power assuming a 12.5~ps pulse duration (middle) and estimated $F_{avg}$ values (bottom) using Eq.~(\ref{equ:2p_abs}) or Eq.~(\ref{equ:2p_abs_sat}) (black line) for the Yb-doped PCF with and without gain at different repetition rates. (b) Anti-Stokes spectra from the Yb-doped PCF at three selected repetition rates.}
\label{fig:anti-Stokes_values}  
\end{figure}

We generated picosecond anti-Stokes pulses centered at 790~nm via FWM in the Yb-doped PCF. The AOM in the 1064~nm fiber MOPA system was used to adjust the repetition rate in discrete steps to identify the optimal setting for TPM. Lowering the repetition rate increased the peak power of the 1064~nm fiber MOPA pulses, which, in turn, enhanced the conversion efficiency of the FWM. As shown in Fig.~\ref{fig:anti-Stokes_values}(a), this improved conversion efficiency initially boosted the average power of the 790~nm pulses. However, at repetition rates $<$7.39~MHz, the lower pulse frequency reduced the average power of the 1064~nm fiber MOPA pulses, which consequently caused a decline in the average power of the 790~nm pulses. Despite this, the peak power of the 790~nm pulses continued to rise [Fig.~\ref{fig:anti-Stokes_values}(a)] due to the higher peak power of the 1064~nm fiber MOPA pulses. At lower repetition rates, the increased peak power of the anti-Stokes pulses resulted in spectral broadening from nonlinear effects such as self-phase modulation [Fig.~\ref{fig:anti-Stokes_values}(b)].

Fig.~\ref{fig:anti-Stokes_values}(a) also highlights the effect of amplification in the Yb-doped PCF. When the 1064~nm pulses were amplified via the Yb-doping in the PCF, significantly higher anti-Stokes average and peak powers were observed at all repetition rates [red points Fig.~\ref{fig:anti-Stokes_values}(a)]. When no 976~nm pump power was applied, there was no population inversion and hence no amplification in the Yb-doped PCF, resulting in lower anti-Stokes average and peak powers [yellow points Fig.~\ref{fig:anti-Stokes_values}(a)].

The temporal characterization of the anti-Stokes pulses was challenging due to their extremely large time-bandwidth product, highly structured optical spectrum and shot-to-shot spectral variations from the noise-seeded nature of the FWM process. These factors rendered commercial pulse characterization devices, based on techniques such as frequency-resolved optical gating (FROG), unsuitable. To estimate the anti-Stokes temporal profile for peak power calculations, we instead simulated the FWM process in the PCF using commercial software that employs the split-step Fourier transform method~\cite{fiberdesk}. To model the noise-seeded nature of FWM, the simulations introduced noise by initializing each frequency bin with a single photon assigned with a random phase. Ten simulations with randomly assigned phases were performed and the anti-Stokes envelope duration was determined by averaging the resulting anti-Stokes pulses. The simulations yielded an estimated anti-Stokes pulse duration of $12.5 \pm 2.5$~ps, which was used for peak power calculations throughout this work.

\subsection{Two-photon microscopy}  

\begin{figure}[htbp]  
\centering\includegraphics{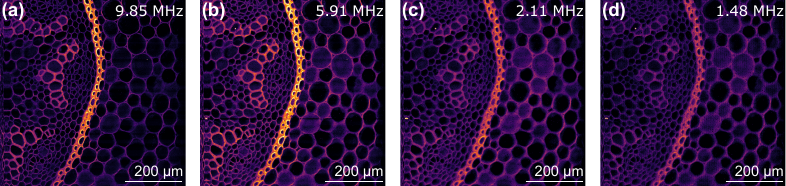}  
\caption{Two-photon excitation fluorescence images of a stained \textit{Convallaria majalis} sample acquired at different repetition rates with the FWM source. The PMT gain and color scale is identical for all images. (a) Highest tested repetition rate (22.9~mW). (b) Repetition rate producing the highest pixel intensities (23.5~mW). (c) Predicted optimal repetition rate from Eq.~(\ref{equ:2p_abs}) (16.7~mW). (d) Lowest tested repetition rate (13.5~mW). Average power on sample in brackets.}  
\label{fig:HF36_lily_rep_rates}  
\end{figure}

To test the two-photon excitation fluorescence (TPEF) imaging performance of our FWM source, we used a histological section of Lily of the valley (\textit{Convallaria majalis}) rhizome stained with acridine orange as the sample. First, we tested the FWM source at different repetition rates to find the optimal value that generates the highest TPEF pixel intensities. This occurred at a repetition rate of 5.91~MHz [Fig.~\ref{fig:HF36_lily_rep_rates}], which was close to the expected value for the maximum time-averaged TPEF flux predicted by Eq.~(\ref{equ:2p_abs_sat}) that takes into account excitation saturation [black line Fig.~\ref{fig:anti-Stokes_values}(a)]. We note that the optimum repetition rate predicted by Eq.~(\ref{equ:2p_abs}) [red points Fig.~\ref{fig:anti-Stokes_values}(a)] is 2.11~MHz, which produced lower TPEF pixel intensities, highlighting the fact that saturation effects are important for this sample and focusing conditions. This was corroborated by the fact the resolution of the images in [Fig.~\ref{fig:HF36_lily_rep_rates}] degrades with decreasing repetition rate, where the pulse energies are higher and thus the effective point spread functions are likely to broaden due to saturation~\cite{cianci2004saturation}. We did not observe photobleaching at any of the tested repetition rates.

To highlight the benefits of amplification in the Yb-doped PCF, we acquired TPEF images with and without amplification by turning the 976~nm pump power on and off, respectively. We optimized the repetition rate to maximize the TPEF intensity for each case, which resulted in different optimal values. For the non-amplified case, the highest TPEF intensity was observed at a repetition rate of 1.48~MHz, which was expected from the predicted TPEF flux using Eq.~(\ref{equ:2p_abs})  [yellow points Fig.~\ref{fig:anti-Stokes_values}(a)], since saturation effects were not significant for these pulse energies. 

\begin{figure}[htbp]  
\centering\includegraphics{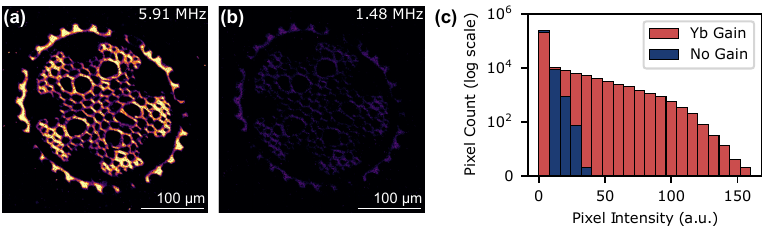}  
\caption{Two-photon excitation fluorescence images of a stained \textit{Convallaria majalis} sample when the Yb-doped PCF was amplifying with 23.5~mW average power on the sample (a) and was not amplifying with 3.5~mW average power on the sample (b). The PMT gain and color scale are identical. (c) Histograms of the image pixel intensities.}
\label{fig:TPI_AFWM_vs_FWM}  
\end{figure}

Fig.~\ref{fig:TPI_AFWM_vs_FWM}(a) shows the TPEF image acquired with amplification in the Yb-doped PCF at 5.91~MHz repetition rate, while Fig.~\ref{fig:TPI_AFWM_vs_FWM}(b) presents the same sample imaged without amplification at a repetition rate of 1.48~MHz. The PMT gain and color scale are the same in both images, highlighting the significant increase in TPEF intensity that results from the amplification in the Yb-doped PCF. This can be observed in the histograms of the pixel intensities, where the amplified case [`Yb Gain' Fig.~\ref{fig:TPI_AFWM_vs_FWM}(c)] has a higher proportions of pixels with greater intensity than the non-amplified case [`No Gain' Fig.~\ref{fig:TPI_AFWM_vs_FWM}(c)]. This confirms that the additional anti-Stokes pulse energy generated through amplification in the Yb-doped PCF increases the TPEF intensity.

\begin{figure}[htbp]  
\centering\includegraphics{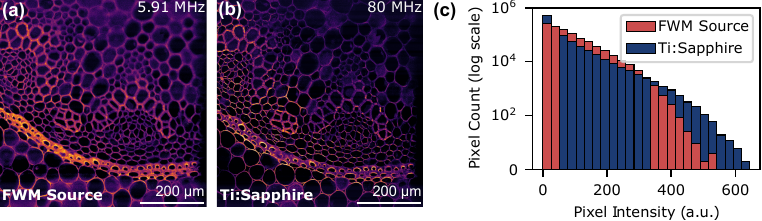}  
\caption{Two-photon excitation fluorescence images of a stained \textit{Convallaria majalis} sample aquired with the FWM source with 23.5~mW average power on the sample (a) and with the Ti:Sapphire laser with 12.0~mW average power on the sample (b). The PMT gain and color scale are identical. (c) Histograms of the image pixel intensities.}
\label{fig:TPI_AFWM_vs_MaiTai}  
\end{figure}

Next, we compared TPEF images acquired with our FWM source against the Ti:Sapphire laser with the average powers adjusted to match the predicted time-averaged TPEF flux using Eq.~(\ref{equ:2p_abs_sat}). The average power of the 80~MHz Ti:Sapphire laser on the sample was set to 12~mW to match the predicted TPEF flux of the FWM source operating at a repetition rate of 5.91~MHz with an average power of 23.5~mW on the sample. As shown in Figs.~\ref{fig:TPI_AFWM_vs_MaiTai}(a) and \ref{fig:TPI_AFWM_vs_MaiTai}(b), our FWM source was able to image the same sample features as the Ti:Sapphire laser with comparable TPEF intensities. Fig.~\ref{fig:TPI_AFWM_vs_MaiTai}(c) shows a histogram of the pixel intensities for the two images. The Ti:Sapphire laser image has a broader range of intensities, with more pixels at both low and high intensities, whereas the FWM source has more pixels at medium intensities.

Fig.~\ref{fig:TPI_AFWM_vs_MaiTai}(b) shows that the Ti:Sapphire laser resolves finer structural details compared with the FWM source at 5.91~MHz repetition rate [Fig.~\ref{fig:TPI_AFWM_vs_MaiTai}(a)]. Performing z-stacks with both lasers revealed that the FWM source imaged more out-of-focus features, suggesting the axial effective point spread function was larger than the Ti:Sapphire laser. We attribute this to excitation saturation, which is corroborated by z-stacks that revealed the depth of field was similar to the Ti:Sapphire laser (80~MHz, \(E_{\mathrm{p}}=0.15\)~nJ)  at the highest tested repetition rate (9.85~MHz, \(E_{\mathrm{p}}=2.3\)~nJ) but increased markedly at the lowest tested repetition rate (1.48~MHz, \(E_{\mathrm{p}}=9.1\)~nJ).

\section{Discussion}
\label{sec:discussion}

Endoscopic TPM for \textit{in vivo} deep tissue imaging\cite{helmchen2001mini, engelbrecht2008ultra, ducourthial2015development, lombardini2018high, septier2024hollow} requires the delivery of high peak power pulses from the laser source through a meters-long fiber to a scanning probe at the distal end. If standard step-index fibers are used, the pulses acquire large group delay dispersion that around 800~nm is dominated by the material dispersion of fused silica (36~fs\(^2\)/mm). Therefore, ultrafast pulses require careful pre-chirping, that also needs to account for the nonlinear chirp of self-phase modulation, to avoid significant temporal broadening at the distal end of the endoscope. Since self-phase modulation ultimately limits the peak power of the pulses that can be delivered through step-index fibers and solid-core PCFs, hollow-core PCFs, where the pulses propagate mainly in air, are commonly used to overcome this limitation~\cite{engelbrecht2008ultra}. Commercial femtosecond fiber lasers are now offered with hollow-core PCF fiber delivery, however, their bulk pulse compressor and free-space hollow-core PCF coupling limit the compactness and robustness of the setup. 

By using picosecond instead of femtosecond excitation pulses, dispersion compensation is not required for either the fiber laser or the fiber delivery. This significantly reduces the cost of the fiber laser (to \(\sim\)\$10k parts cost) and allows full fiber integration, enabling highly compact, robust endoscopic TPM devices to be realized. In principle, the same time-averaged TPEF flux can be generated using picosecond pulses compared with femtosecond pulses at fixed average power if the laser repetition rate is reduced proportionally according to Eq.~(\ref{equ:2p_abs})~\cite{karpf2016two, kunio2024efficient}. However, we have found when imaging a histological stained \textit{Convallaria majalis} sample using a 20x 0.5~NA objective that excitation saturation effects become important at low repetition rates where the pulse energy is high. When saturation occurs, the TPEF flux generated does not continue to increase as the repetition rate is reduced at fixed average power, as would be expected from Eq.~(\ref{equ:2p_abs}). Instead, there is an optimum repetition rate that maximizes the TPEF flux generated for a given average power and pulse duration, as described by Eq.~(\ref{equ:2p_abs_sat}).

The variable repetition rate of our 12.5~ps FWM source enabled us to determine that the optimal repetition rate that generated the highest TPEF pixel intensities for our sample and focusing conditions was 5.91~MHz, corresponding to a pulse energy of 4~nJ on the sample. Some degradation of the resolution was observed at this repetition rate, due to broadening of the radial and axial effective point spread functions from saturation~\cite{cianci2004saturation}. Operating at the highest tested repetition rate of 9.85~MHz reduced this resolution degradation with a marginal reduction in the TPEF pixel intensities. The repetition rate flexibility of our FWM source enables the pulse energy to be tailored to the amount of saturation desired and the saturation energy of the fluorophore, which is dependent on the two-photon cross section and the focusing conditions.

The balance of pulse energy and saturation effects has previously been discussed in the context of deep tissue imaging with a variable repetition rate ultrafast fiber laser source~\cite{charan2018fiber}. We have only considered thin samples in this work and hence the scaling of the optimum repetition rate in Eq.~(\ref{equ:opt_rr}) has no dependence on the imaging depth relative to the attenuation length. The variable repetition rate of our FWM source over a similar frequency range could likewise be useful for optimizing the TPEF signal at different imaging depths. However, in cases where linear thermal damage is the limiting factor, shorter pulse durations would be preferred to maximize the TPEF signal. Compared with~\cite{charan2018fiber}, our FWM source is simpler, less expensive and has the potential to be fully fiber integrated, whilst also eliminating the need for any dispersion compensation.

\section{Conclusion}  

Femtosecond solid-state lasers are the predominant sources for two-photon microscopy (TPM), providing pulses with high peak power but at the expense of bulkiness, high cost of ownership and limited portability. In this work, we demonstrated a compact and cost-effective alternative by using four-wave mixing (FWM) in a Yb-doped photonic crystal fiber (PCF) to generate picosecond anti-Stokes pulses at 790~nm. Yb-doping in the PCF played a crucial role by amplifying the 1064~nm input pulses, significantly enhancing the anti-Stokes average and peak power generated. The fiber-based picosecond FWM source was benchmarked against a femtosecond Ti:Sapphire laser using a commercial two-photon microscope. Despite its longer pulse duration, our source produced comparable two-photon excitation fluorescence intensity and image quality under matched two-photon excitation conditions. The Yb-doped PCF can be fully fiber-integrated to create an alignment-free, compact picosecond source that is suitable for endoscopic TPM.

\begin{backmatter}

\bmsection{Funding} Royal Academy of Engineering (RF/201920/19/285); Engineering and Physical Sciences Research Council (EP/W024020/1).

\bmsection{Acknowledgments} BK acknowledges support from an EPSRC funded studentship. THR is supported by the Royal Academy of Engineering under the Research Fellowship scheme.

\bmsection{Disclosures} 
The authors declare no conflicts of interest.

\bmsection{Data Availability} 
Data underlying the results presented in this paper may be obtained from the authors upon reasonable request.
\end{backmatter}

\newpage
\bibliography{Bibliography}

\end{document}